\documentclass{article}

\PassOptionsToPackage{numbers}{natbib}

 \usepackage[main, final]{neurips_2026}

\usepackage[utf8]{inputenc} 
\usepackage[T1]{fontenc}    
\usepackage[hidelinks]{hyperref}       
\usepackage{url}            
\usepackage{booktabs}       
\usepackage{amsfonts}       
\usepackage{nicefrac}       
\usepackage{microtype}      
\usepackage{xcolor}         

\title{Privacy in Personalized AI Is a System Property, Not~Just~a~Model~Property}

\author{%
  Guillaume Salha-Galvan \\
  SJTU Paris Elite Institute of Technology\\
  \texttt{gsalhagalvan@sjtu.edu.cn} \\
   \And
   Jiaying Xu \\
 Kibo Ryoku Research \\
 \texttt{jxu@kiboryoku.com}
}

\begin{document}

\maketitle

\begin{abstract}
In personalized AI applications, such as conversational assistants and recommender systems, users interact not with models in isolation but with broader systems that access, infer, and reuse user information across components and over time. While such use of user information is integral to personalization, it also raises important privacy questions. In this paper, we argue that individual model- or component-level analyses may not capture all privacy risks arising in such systems, motivating a system-level perspective on privacy. We distinguish and analyze four interconnected privacy-risk channels in personalized AI, and subsequently propose four requirements for system-level privacy evaluation, covering interaction trajectories, internal information flows, indirect leakage, and the privacy--utility trade-off. We argue for their systematic incorporation into privacy audits of personalized AI.
\end{abstract}

\section{Introduction}

Privacy has long been a central concern in machine learning \cite{abadi2016deep,dwork2006calibrating,el2024preserving,pan2024differential,shokri2017membership}, and its importance~has only increased in the era of foundation models \cite{bommasani2021opportunities,das2025security,kibriya2024privacy,miranda2025preserving,wang2025unique}. As models become larger,~more opaque, and more deeply embedded in real-world applications, privacy risks are increasingly difficult to characterize and mitigate. Substantial efforts across artificial intelligence (AI) research and practice continue to study how models may expose or compromise user information~\cite{carlini2021extracting,carlini2023quantifying,lukas2023analyzing,mattern2023membership,nasr2023scalable,staab2024beyond,wang2025unveiling}.

However, this model-centric perspective becomes more difficult to apply when considering privacy in personalized AI applications, such as conversational assistants or recommender systems \cite{briand2021semi,gao2021advances,zhang2025personalization}. In these settings, users interact not with isolated models but with broader systems built around them, combining core models with user profiles, interaction histories, memories, retrieval mechanisms including retrieval-augmented generation, ranking mechanisms, external tools, or agentic pipelines \cite{delcluze2026music,jiang2025know,lewis2020retrieval,wang2024crafting,yuan2025personalized,zhang2026idproxy}. Accessing user information is integral to the purpose of personalization, but~such information may be used, inferred, and exposed across multiple components and over time, sometimes implicitly or indirectly \cite{huang2025recommender,maragheh2025future,mireshghallah2024can,mukhopadhyay2025privacybench}. This raises a fundamental question: is analyzing the privacy of individual models or components sufficient to characterize~privacy~in~personalized~AI?

Consider, for example, a personalized AI assistant helping a user organize their daily life. Even if its underlying language model was never trained on the user's private data, the assistant may access calendars, past conversations, emails, or other connected tools. When asked \textit{"What should I do this morning?"}, it might mention a sensitive medical appointment or reveal private information from earlier conversations \cite{mireshghallah2024can,mukhopadhyay2025privacybench,zharmagambetov2025agentdam}. Similarly, a recommender system personalizing a social media feed may infer sensitive preferences, such as political interests, from user behavior accumulated over time and expose them through its recommendations, even if the user never explicitly disclosed them. An external observer could then inappropriately learn about these~preferences \cite{beigi2020privacy,volkova2014inferring,xin2023user}.

In practice, such risks are still often studied through analyses targeting individual models, components, or leakage mechanisms, including memorization in foundation models, leakage from persistent memory, inappropriate information retrieval, or privacy attacks on recommender systems \cite{
beigi2020privacy,
carlini2023quantifying,
mireshghallah2024can,
mukhopadhyay2025privacybench,
wang2025unveiling,
xin2023user,
zhang2026privacypeek,
zhang2023comprehensive,
zharmagambetov2025agentdam}. While these analyses remain important, we argue in this paper that treating these risks separately can obscure a broader issue. Our position is that privacy in personalized AI should be treated as a property of the system as a whole, with the full user--system interaction and its information flows across components and over time as the unit of analysis.

To develop this position, we distinguish and analyze four interconnected privacy-risk channels in personalized AI: data-access, inferential, behavioral, and compositional leakage. Building on this framework, we then propose four requirements to guide system-level privacy evaluation in future privacy audits, covering interaction trajectories, information flows, indirect leakage, and the privacy--utility trade-off. We finally discuss the implications and boundaries of our position~and~requirements.

\section{Personalization Expands the Privacy Surface}
\label{section2}

\paragraph{Personalized AI Systems}
Personalized AI is increasingly embedded in everyday digital services, from conversational AI assistants tailored to individual users to recommender systems deployed across social media, e-commerce, and streaming platforms, where they help users navigate large catalogs and personalize their online experience \cite{bendada2023track,bendada2023scalable,li2024recent,zhang2026idproxy,zhang2025personalization}. Personalization expands the privacy surface in a specific way: system behavior depends on information collected, inferred, and reused about users. Unlike a standalone model receiving an isolated input, a personalized AI system may accumulate information across interactions, maintain explicit or implicit user representations, retrieve information when needed, route information through external tools, and adapt its outputs accordingly \cite{huang2025recommender,jiang2025know,yuan2025personalized,zhang2025personalization}. Privacy risks may therefore arise not only from information directly observed and stored by a model, but also from what the system accesses, infers, exposes through its behavior, and combines over time.

\paragraph{Privacy-Risk Channels}
In this paper, we distinguish four complementary and interconnected privacy-risk channels in personalized AI. These channels are not mutually exclusive, but rather aim to provide a conceptual framework for characterizing how privacy loss may manifest in this setting:
\begin{itemize}

\item \textbf{Data-access leakage:} private information available to the system may be accessed or transmitted beyond what is necessary, expected, or authorized for the current task. This information may come from a user profile, interaction history, retrieved document, database, or connected service. For example, an AI assistant recommending a restaurant may retrieve medical information to account for allergies, even though the user may not expect or be comfortable with such sensitive information being accessed for this purpose. Similarly, a shopping agent may inappropriately transmit a user's home address to an external tool, even if it never appears in the user-facing output. Recent evaluations of personalized assistants and tool-using agents have indeed documented cases involving indiscriminate retrieval of sensitive information and unnecessary data acquisition during task execution~\cite{mukhopadhyay2025privacybench,zhang2026privacypeek,zharmagambetov2025agentdam}.

\item \textbf{Inferential leakage:} a personalized system may also infer sensitive information. For example, a social media recommender system may combine viewing and liking patterns to infer a user's political interests even if the user never declared them. Similarly, a shopping assistant may infer a sensitive life event from searches or purchases. This is not inherently a privacy violation: personalization inherently relies on inferring user preferences and intentions. It becomes privacy-relevant when the system derives sensitive knowledge beyond what is necessary or reasonably expected for the intended task, particularly when that knowledge is retained, reused across contexts, or exploited against the user's interests or expectations. For instance, inferred political preferences could be reused for personalization even though the user never chose to disclose them or consent to such use \cite{mireshghallah2025position,staab2024beyond,volkova2014inferring,zhang2023comprehensive}.

\item \textbf{Behavioral leakage:} regardless of whether private information is explicitly stored or internally inferred, it may become observable to others through the system's personalized behavior. Behavioral leakage concerns what others can learn from personalized outputs or actions, rather than what the system itself infers. Rankings, recommendations, explanations, advertisements, or other actions may all expose private characteristics of the user. For example, a social media feed prominently recommending content associated with a political movement may expose the user's political interests. Similarly, a personalized assistant explaining why it made a recommendation may inadvertently reveal information from the user's private history. Prior research has shown that recommendation outputs can be exploited to infer private attributes or reconstruct user behavior~\cite{beigi2020privacy,calandrino2011you,xin2023user}.

\item \textbf{Compositional leakage:} finally, privacy loss may emerge only when multiple components, contexts, or interactions are considered together, even if none appears particularly revealing in isolation. An observer may repeatedly probe an assistant, compare personalized outputs over time, correlate information across contexts, or combine signals exchanged between memories, tools, and agents. For example, a calendar entry, an earlier conversation, and subsequent recommendations may each appear innocuous on their own, yet jointly reveal that a user is undergoing medical treatment. Similarly, information legitimately accessed for one purpose may become privacy-sensitive when reused in another context: a health-related detail shared with a personal assistant, for instance, might later influence recommendations in an unrelated shopping or entertainment service, exposing information that the user never intended to carry across contexts. Compositional leakage therefore concerns privacy risks that become visible only at the level of an interaction trajectory as a whole, rather than through any single response or component \cite{debenedetti2024privacy,el2026agentleak,mireshghallah2026cimemories,patil2025sum}.

\end{itemize}

\paragraph{Discussion}
This framework complements rather than replaces model-level privacy analyses. Training-data memorization, extraction, and related attacks remain important and may coexist with the risks considered here \cite{carlini2021extracting,carlini2023quantifying,mattern2023membership,shokri2017membership}. Our position is that model-level analyses alone do not capture all privacy risks introduced or amplified when a model becomes part of a personalized system. Moreover, our four channels aim to describe how privacy loss manifests, rather than the mechanisms that trigger it. Prompt injection, tool misuse, or adversarial attacks, for instance, may activate one or more leakage channels rather than constituting separate channels~themselves~\cite{debenedetti2024agentdojo,fu2024imprompter,he2025emerged}.

Finally, the four channels deliberately overlap. Consider a private medical event retrieved from a user's calendar. Its retrieval may constitute data-access leakage; the system may use it to infer a health-related attribute; this inference may alter subsequent recommendations and create behavioral leakage; and repeated observations of those recommendations may expose further information through compositional leakage. No single component or output needs to reveal the full privacy-sensitive picture. It is precisely this interaction between access, inference, behavior, and composition that expands the privacy surface of personalized AI and motivates a system-level perspective on privacy.

\section{From Model Audits to System Audits}

\paragraph{Requirements for System-Level Privacy Evaluation} Our position implies a shift toward system-level privacy evaluation. Building on the framework of Section~\ref{section2}, we propose four requirements and advocate for their systematic incorporation into future privacy audits of personalized AI applications:

\begin{itemize}

\item \textbf{Evaluate interaction trajectories, not only individual responses:} as personalized systems accumulate information and adapt over time, privacy risks may emerge only across multiple interactions (compositional leakage). Privacy evaluations should therefore include longitudinal user histories and allow adaptive probing based on prior outputs. Several individually innocuous responses may jointly reveal sensitive information that no single response exposes \cite{patil2025sum}. Recent benchmarks such as PrivacyBench and CIMemories \cite{mireshghallah2026cimemories,mukhopadhyay2025privacybench} have begun moving in this direction by evaluating privacy over multi-turn conversations, persistent user memories, and repeated interactions. We argue that longitudinal evaluation should become standard in privacy audits of personalized AI.

\item \textbf{Audit internal information flows, not only final outputs:} output-only evaluation may miss privacy risks occurring inside the system (data-access leakage and, potentially, compositional leakage). An assistant may retrieve or transmit private information beyond what is appropriate without revealing it in its final response. Privacy audits should therefore examine, when accessible, which information is retrieved, which tools or agents receive sensitive data, and what is retained afterward. Such access is not necessarily a privacy violation; the key question is whether it is necessary, authorized, and appropriate for the current task and context. Recent benchmarks such as AgentLeak, PrivacyPeek, and ToolPrivacyBench have moved beyond output-only evaluation by auditing sensitive-data acquisition, internal communication channels, retrieval behavior, and information flows through tool calls \cite{el2026agentleak,hu2026toolprivacybench,zhang2026privacypeek}. We argue that privacy audits of personalized AI should systematically account for such hidden information flows rather than focus only~on~user-facing~outputs.

\item \textbf{Do not neglect inferential and behavioral leakage:} privacy evaluations focused only on predefined secrets or explicit disclosure may fail to characterize leakage of private information that personalization derives or exposes indirectly (inferential and behavioral leakage). Audits should therefore evaluate what sensitive attributes the system can infer from user behavior, as well as what an external observer can learn about the user from the system's personalized outputs or actions, such as recommendation feeds \cite{beigi2020privacy,calandrino2011you,volkova2014inferring,xin2023user}. Both forms of leakage should be systematically considered when evaluating privacy~in~personalized~AI.

\item \textbf{Assess privacy jointly with personalization utility:} preventing a system from accessing or inferring any user information would reduce personalization-related privacy risks while defeating the purpose of personalization. Conversely, unrestricted access to user data may improve personalization while increasing risks across the four channels from Section~\ref{section2}. Privacy evaluation should therefore assess privacy and utility jointly rather than in isolation. For example, a travel assistant may need access to a user's destination and preferences to provide useful recommendations, but not to unrelated medical information or long conversation history. Audits should consider task utility alongside unnecessary information access, sensitive inference, behavioral disclosure, and privacy loss over repeated interactions. This also connects to data minimization: the relevant question is not only whether information is protected, but whether the system needs to access or retain it in~the~first~place~\cite{biega2020operationalizing,zharmagambetov2025agentdam}.
\end{itemize}

\paragraph{Discussion}
These requirements operationalize the system-level perspective developed in Section~\ref{section2}. We next clarify several implications and boundaries of our position. First, we do not argue that every system should be evaluated identically. On the contrary, different applications involve different components, information flows, and forms of personalization. Privacy audits should therefore make their threat model explicit, specifying which user data the system is authorized to access or infer, for what purpose, and which components or observers may receive sensitive information \cite{barth2006privacy,liao2026linddun,mireshghallah2024can}. 

Second, we stress that our position is more specific than the general observation that privacy should be considered end-to-end across a deployed system \cite{debenedetti2024privacy}. We argue that personalization creates interconnected privacy-risk channels across components and over time. Even if every individual component passes its own privacy evaluation, and no privacy failure is detected through isolated input-output evaluation, the personalized system as a whole may still expose privacy risks arising from dependencies among components and across successive interactions \cite{el2026agentleak,patil2025sum}. The four channels identified in Section~\ref{section2} are intended to capture precisely such system-level risks.

Third, we do not aim to diminish the importance of model-level analysis or formal privacy guarantees. Differential privacy, for example, can provide strong guarantees limiting the influence of individual training examples \cite{abadi2016deep,dwork2006calibrating}. Such guarantees remain valuable, but they generally apply within a particular threat model and to a model, mechanism, or processing stage \cite{debenedetti2024privacy}. They do not characterize risks arising when sensitive user information is later introduced through profiles, memories, retrieval mechanisms, or external tools \cite{el2026agentleak,mukhopadhyay2025privacybench,zhang2026privacypeek}. In contrast, the system-level audits we advocate do not provide formal privacy guarantees, but complement formal analyses by capturing risks beyond their scope~\cite{debenedetti2024privacy}. We view model- and system-level privacy analyses as complementary~rather~than~competing.

Fourth, as discussed in this paper, recognizing inferential leakage as privacy-relevant does not imply that inference itself is undesirable. Inferring user preferences and needs is often an integral part of personalization. Rather, the relevant question is whether sensitive inferences are necessary or reasonably expected for the intended purpose, and how they are subsequently retained, reused,~or~exposed.

Finally, a system-level perspective calls for precision in privacy claims \cite{brown2022does,debenedetti2024privacy}. For example, \textit{"the model is differentially private," "the model did not memorize the user's data,"} and \textit{"the system protects the user's privacy"} are three distinct statements. The first two concern models or mechanisms; the last concerns the personalized interaction and surrounding information flows. Conflating these levels risks turning component-level guarantees into unsupported system-level privacy claims.

\section{Conclusion}
We argued that privacy in modern personalized AI applications cannot be reduced to a model property. We showed that personalization expands the privacy surface through four interconnected channels: data-access, inferential, behavioral, and compositional leakage. Building on this framework, we proposed four requirements for system-level privacy evaluation, covering interaction trajectories, internal information flows, indirect leakage, and the privacy--utility trade-off. Together, these contributions aim to provide a cohesive basis for system-level privacy evaluation in personalized AI.

\bibliographystyle{plainnat} 
\bibliography{references}

\end{document}